\documentclass[aps,prb,twocolumn,showpacs,amsmath,amssymb,unsortedaddress]{revtex4-1}
\usepackage{graphicx}
\usepackage{longtable}
\usepackage{ulem}
\usepackage{color}

\begin{document}
\title{The electronic structure of Co-substituted $\textrm{Fe}\textrm{Se}$ superconductor probed by soft X-ray spectroscopy and density functional theory}

\author{I. Perez} 
\email[Contact Author: ]{cooguion@yahoo.com}
\affiliation{Department of Physics and Engineering Physics, University of Saskatchewan, Saskatoon SK Canada, 116 Science Place, S7N 5E2}
\author{J. A. McLeod} 
\affiliation{Department of Physics and Engineering Physics, University of Saskatchewan, Saskatoon SK Canada, 116 Science Place, S7N 5E2}
\author{R. J. Green} 
\affiliation{Department of Physics and Engineering Physics, University of Saskatchewan, Saskatoon SK Canada, 116 Science Place, S7N 5E2}
\author{R. Escamilla}
\affiliation{{\it Instituto de Investigaciones en Materiales, Universidad Nacional Aut\'onoma de M\'exico, C. P. 04510 M\'exico D.F.}}
\author{V. Ortiz}
\affiliation{{\it Instituto de Investigaciones en Materiales, Universidad Nacional Aut\'onoma de M\'exico, C. P. 04510 M\'exico D.F.}}
\author{A. Moewes} 
\affiliation{Department of Physics and Engineering Physics, University of Saskatchewan, Saskatoon SK Canada, 116 Science Place, S7N 5E2}
\date{\today}

\begin{abstract}
We study the crystalline and electronic properties of the $\textrm{Fe}_{1-x}\textrm{Co}_x\textrm{Se}$ system ($x=0$, 0.25, 0.5, 0.75, and 1.0) using X-ray diffraction, X-ray spectroscopy and density functional theory. We show that the introduction of Co $3d$ states in FeSe relaxes the bond strengths and induces a structural transition from tetragonal to hexagonal whose crossover takes place at $x\approx0.38$. This structural transition in turn modifies the magnetic order which can be related to the spin state. Using resonant inelastic X-ray spectroscopy we estimate the spin state of the system; FeSe is found to be in a high spin state (S=2), but Fe is reduced to a low spin state upon Co substitution of $x \le 0.25$, well below the structural transition. Finally, we show evidence that FeSe is a moderately correlated system but the introduction of Co into the host lattice weakens the correlation strength for $x\ge0.25$. These novel findings are important to unravel the mechanisms responsible for the superconducting state in iron-chalcogenide superconductors. 

\end{abstract}
\maketitle
\section{Introduction}

In 2008, a breakthrough in condensed matter physics was announced by Japanese researchers \cite{kamihara} --- a new family of superconductors based on the element iron was discovered with relatively high critical temperatures ($T_c$). \cite{rotter,wang5} Among the new materials, the iron chalcogenides with $T_c$'s ranging from 7 K to 15 K at normal pressure have been the focus of intense research during the last years due to their structural simplicity and the absence of toxic elements, features that make them potential candidates for industrial applications.\cite{hsu,ma} In 2010, the Fe-Se layer compound K$_{0.8}$Fe$_2$Se$_2$ with a relatively high $T_c$ of 31 K was discovered.\cite{guo10b} This new discovery intensified the interest in understanding the underlying mechanisms responsible for the superconducting state in these materials. 

The iron-chalcogenide superconductor $\alpha$-FeSe has a tetragonal PbO-type structure with space group $P4/nmm$ at room temperature and undergoes an orthorhombic structural transition between 70 K and 90 K.\cite{margadonna, mcqueen} Its superconducting state can be manipulated by applying hydrostatic or chemical pressure. Enhancements in the $T_c$ of up to 37 K have been attained applying pressures of 7 GPa.\cite{gresty,kumar} On the other hand, investigations on the effect of Te substitution on the superconducting and crystalline properties of FeSe show that the $T_c$ reaches a maximum of 15 K for Te substitution of 50\% while the lattice parameters increase proportionally to Te concentration. These cases highlight the influence of the crystalline properties on the electronic structure of these materials. The effect of chalcogen-substitution on the electronic properties of FeSe has been widely explored in the literature. \cite{wu4,deguchi,fang,kawashima,yeh,yeh1,viennois1,kurmaev,simonelli,subedi} X-ray emission spectroscopy (XES) and resonant inelastic X-ray scattering (RIXS) measurements performed at the Fe $L_{2,3}$ edge of the $\textrm{FeSe}_{1-x}\textrm{Te}_{x}$ system show that the spin state varies between 0 and 2 as function of $x$ with Fe in FeSe in the highest spin state.\cite{iperez14b} Surprisingly, Te substitution has a minimal effect on the density of states that persists for Te substitution between $x=0$ and $x=0.75$. The measurements also unveil the itinerant character of the $3d$ electrons and strongly suggest that the $\textrm{FeSe}_{1-x}\textrm{Te}_{x}$ system can be regarded, at most, as an intermediate correlated system, similar to the case of iron-pnictides,\cite{kurmaev1, kurmaev2, mcleod5, singh,singh1,yang1} although the possibility of FeSe being a strongly correlated system has not been ruled out yet.\cite{miyake10a,aichhorn1,yamasaki10a} The electronic structure of the $\textrm{FeS}_{1-x}\textrm{Te}_{x}$ system was also investigated with extended X-ray absorption fine structure measurements.\cite{iadecola11a} The authors reported that the Fe-Te and Fe-S bond lengths are inequivalent and that this system should be treated as random alloy. 

Although great advances have been made, the role the Fe valence states play in order to favour the superconducting state in FeSe is still not well understood. A natural path that could shed light on this issue is to substitute the Fe site by a transition metal (TM=Co, Ni or Cu) and investigate its effects on the superconducting and electronic properties. The influence on the electronic band structure of Fe-pnictides due to doping/substitution with TM atoms has been explored both theoretically and experimentally albeit the picture that emerged is unclear. Two scenarios are being discussed: (1) Theoretical work based on first-principles calculations via density functional theory (DFT) suggests that TM doping is isovalent with Fe and does not change the amount of free charge carriers in the system. As a consequence, the dopants do not make great contributions to the density of states at the Fermi energy ($E_F$), although, the Fermi surface topology could be considerably influenced in such a way that the size of the electron (hole) pockets at the Brillouin zone is modified to enhance the superconducting state.\cite{wadati} This view is supported by X-ray absorption spectroscopy experiments \cite{merz} performed at the Fe and Co $L$ edges of $\textrm{Sr}(\textrm{Fe}_{1-x}\textrm{Co}_{x})_2\textrm{As}_2$. (2) On the other hand, experience with cuprates teaches us that doping provides additional charge carriers to the system that are essential for driving the material from the spin density wave phase (or the antiferromagnetic phase) to the superconducting one. Likewise, some researchers believe that the presence of extra electrons at the Fermi level may be crucial for the emergence of superconductivity in Fe-pnictide superconductors.\cite{liu3,levy,berlijn,mizuguchi09a} Early soft X-ray spectroscopic measurements \cite{mcleod5,kroll08a} performed on electron doped-materials such as $\textrm{Ba}(\textrm{Fe}_{0.95}\textrm{Co}_{0.5})_2\textrm{As}_2$ and $\textrm{LaFeAsO}_{1-x}\textrm{F}_x$ support this scenario and, indeed, an increase in the carrier concentration in the $\textrm{Ba}(\textrm{Fe}_{x}\textrm{Co}_{1-x})_2\textrm{As}_2$ system as function of $x$ was reported from Hall measurements.\cite{fang1,sefat1} Some calculations \cite{wadati,ciechan,ding,chadov} carried out to investigate the electronic properties in TM-doped iron chalcogenides appear to follow scenario (1). This idea has found support from very recent work \cite{zhang11a,zhang} reporting on the magnetic and superconducting properties of $\textrm{Fe}_{1+y}\textrm{Se}_{1-x}\textrm{Te}_{x}$ doped with TM, albeit, due to the nature of the measurements, no specific information about the dopant electronic structure has been provided.

A systematic experimental study is necessary not only to verify theoretical models but also to provide more information on the electronic properties, such as the spin state and the correlation strength, that have not been explored for TM-substituted FeSe superconductor. RIXS and XES provide element and symmetry specific probes of the partial occupied states and therefore are appropriate tools to study the electronic structure of these materials. In this paper, we perform RIXS and XES measurements for the system $\textrm{Fe}_{1-x}\textrm{Co}_x\textrm{Se}$ ($x=0-1$) and support our measurements with DFT calculations. Using these tools, we discover that the introduction of Co injects free electrons into the system inducing transitions in the electronic correlation strength and the spin state of the system. We find that electrons in FeSe are more localized than in the rest of the series and that FeSe is in a high spin state whereas the rest of the series is in a low spin state. We think that these findings will be important for understanding the pairing mechanisms responsible for superconductivity in these materials.

\section{Experiments and Calculations}
\subsection{Synthesis and crystalline structure of $\textrm{Fe}_{1-x}\textrm{Co}_x\textrm{Se}$}
Samples for the $\textrm{Fe}_{1-x}\textrm{Co}_x\textrm{Se}$ system  with stoichiometries $x=0.00$, 0.25, 0.50, 0.75 and 1.00 were synthesized from Alfa Aesar powders of Fe Putatronic 99.998\%, Co 99.9\% and Se 99.5\% with the required stoichiometry to get 1.50 g samples. The reagents were mixed at the optimal quantities in an agate mortar, then the resulting mixture was introduced and sealed in an evacuated quartz tube at $10^{-2}$ Torr. The tubes were placed in a furnace at 750 $^\circ$C for 7 days and the system was allowed to cool down to room temperature. Finally, powders were ground again with a ball mill to get fine powders and the resulting samples were kept in polyethylene containers. 

\begin{figure}[tp]
\begin{center}
\includegraphics[width=7cm,angle=270]{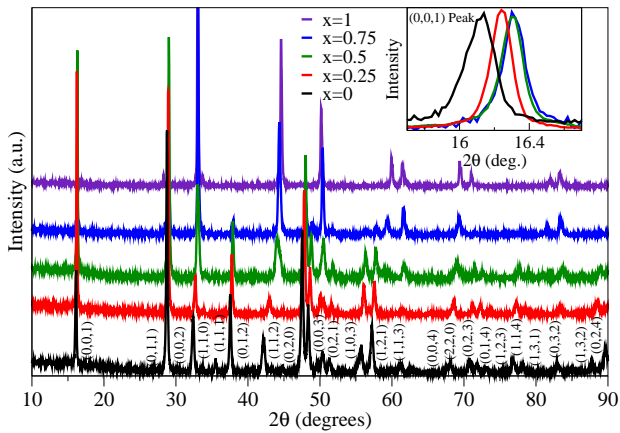}
\caption{(Color online) X-ray diffraction patterns for the system $\textrm{Fe}_{1-x}\textrm{Co}_x\textrm{Se}$ with $x=0-1$. Inset shows a zoom of the (0,0,1) peak as $x$ varies. Note how the peak shifts toward high angles with increasing $x$.}
\label{xrdfecose}
\end{center}
\end{figure}
The crystalline structure of the samples was determined with a Siemens D5000 X-ray diffractometer using a cobalt filament and an iron filter. Operation parameters were set to 34 kV and 30 mA. Phase identification was performed using the ICDD 2004 database. Intensities were measured at room temperature in 0.02$^\circ$ steps, in the 6 degrees - 130 degrees 2$\theta$ range. The Rietveld refinement was carried out using the program MAUD v2.33.

The X-ray diffraction patterns of $\textrm{Fe}_{1-x}\textrm{Co}_x\textrm{Se}$ are displayed in Fig. \ref{xrdfecose}. The inset shows the evolution of the (0,0,1) peak as the $x$ increases. This behaviour is reflected in the variation of the lattice parameters (see Table \ref{fecosetable}); $a$ decreases as $x$ increases, whereas $c$ seems to reach a saturation value between $x=0.5$ and 0.75 indicating a probable solid solubility limit of Co. \cite{mizuguchi09a,liu1} For Co substitution of less than $50\%$, the value of the lattice parameters $a$ and $c$ decreases similarly to Co-substituted Fe-pnictides.\cite{sefat1,wang7, sefat} We point out, however, that Co substitutions of more than 38\% have not been reported in the literature for most iron-based superconductors; most probably because the superconductivity is suppressed for $x>0.38$. This trend may suggest that the size of the lattice parameters decreases monotonically and that the same structural phase is conserved as $x$ varies. However, from Table \ref{fecosetable} it is clear that occurs a structural transition from tetragonal to hexagonal (space group $P6_3/mmc$) as Co content increases. 
 \begin{table}[tp]
   \caption{Structural parameters for the samples used herein.}
  \centering 
  \begin{tabular}{c|ccccc}
  \hline   \hline
$x$ & \footnotesize{0} & \footnotesize{0.25} &  \footnotesize{0.5} & \footnotesize{0.75} &  \footnotesize{1}
  \\ \hline   \hline
\footnotesize{a(\AA)} &   \footnotesize{3.773} & \footnotesize{3.744} & \footnotesize{3.731} & \footnotesize{3.730}  &  \footnotesize{3.620} \\ 
\footnotesize{c(\AA)} &   \footnotesize{5.521} & \footnotesize{5.465} & \footnotesize{5.442} & \footnotesize{5.450}  &  \footnotesize{5.291} \\ 
\footnotesize{V(\AA$^3$)} &   \footnotesize{78.63} & \footnotesize{76.45} & \footnotesize{75.78} & \footnotesize{75.84}  &  \footnotesize{60.08}
  \\ \hline
  \footnotesize{\%$\textrm{Fe}_{1-x}\textrm{Co}_x\textrm{Se}_{0.92}$}  ($P4/nmm$)&   \footnotesize{86.02} & \footnotesize{81.71} & \footnotesize{26.93} & \footnotesize{25.94}  &  \footnotesize{-} \\ 
\footnotesize{\%$\textrm{Fe}_{1-x}\textrm{Co}_x\textrm{Se}$}  ($P6_3/mmc$)&   \footnotesize{-} & \footnotesize{18.29} & \footnotesize{73.07} & \footnotesize{74.07}  &  \footnotesize{93.7} \\ 
\footnotesize{\%$\textrm{Fe}_{7}\textrm{Se}_{8}$} ($P3_121$) &   \footnotesize{13.79} & \footnotesize{-} & \footnotesize{-} & \footnotesize{-}  &  \footnotesize{-} \\ 
\footnotesize{\%$\textrm{Fe}_{3}\textrm{O}_4$}  ($Fd$-$3m$)&   \footnotesize{0.19} & \footnotesize{-} & \footnotesize{-} & \footnotesize{-}  &  \footnotesize{-} \\ 
\footnotesize{\%$\textrm{Co}$}  ($P6_3/mmc$)&   \footnotesize{-} & \footnotesize{-} & \footnotesize{-} & \footnotesize{-}  &  \footnotesize{6.7}\\ \hline
  
    \footnotesize{Bond length (\AA)} &  &  &  &  &   \\ 
  \footnotesize{(Fe/Co)-(Fe/Co):4} &  \footnotesize{2.669} & \footnotesize{2.653} & \footnotesize{2.638} & \footnotesize{2.637}  &  \footnotesize{3.620} \\ 
  \footnotesize{(Se)-(Fe/Co):4} & \footnotesize{2.371} & \footnotesize{2.356} & \footnotesize{2.341} & \footnotesize{2.342}  &  \footnotesize{2.474} \\   \hline
  
    \footnotesize{Bond angle ($^o$)} &  \footnotesize{} & \footnotesize{} & \footnotesize{} & \footnotesize{}  &  \footnotesize{} \\ 
  \footnotesize{Se-(Fe/Co)-Se} &  \footnotesize{105.48} & \footnotesize{105.57} & \footnotesize{105.64} & \footnotesize{105.55}  &  \footnotesize{55.65} \\ 
  \footnotesize{Se-(Fe/Co)-Se} & \footnotesize{68.49} & \footnotesize{68.54} & \footnotesize{68.51} & \footnotesize{68.53}  &  \footnotesize{94.08} \\  \hline

    \footnotesize{N} &  &  &  &   &    \\
    \footnotesize{Fe} & \footnotesize{1.00} & \footnotesize{0.75} & \footnotesize{0.5} & \footnotesize{0.25}  &  \footnotesize{0} \\  
      \footnotesize{Co} & \footnotesize{0} & \footnotesize{0.25} & \footnotesize{0.5} & \footnotesize{0.75}  &  \footnotesize{1.00} \\  
            \footnotesize{Se} & \footnotesize{0.94} & \footnotesize{0.92} & \footnotesize{0.97} & \footnotesize{1.00}  &  \footnotesize{1.00} \\  \hline
      
        \footnotesize{$R_{wp}(\%)$} & \footnotesize{12.95} & \footnotesize{17.10} & \footnotesize{14.02} & \footnotesize{14.27}  &  \footnotesize{14.25} \\  
             \footnotesize{$R_{p}(\%)$} & \footnotesize{9.75} & \footnotesize{13.51} & \footnotesize{10.82} & \footnotesize{10.13}  &  \footnotesize{10.99} \\  
             \footnotesize{$R_{exp}(\%)$} & \footnotesize{5.64} & \footnotesize{9.21} & \footnotesize{7.76} & \footnotesize{8.25}  &  \footnotesize{7.43} \\  
          \footnotesize{$\chi^2$} & \footnotesize{2.29} & \footnotesize{1.86} & \footnotesize{1.81} & \footnotesize{1.73}  &  \footnotesize{1.91} \\  \hline   \hline
\end{tabular}
\label{fecosetable}
\end{table}
Note that the hexagonal phase is not present in FeSe ($x=0$) albeit some other work\cite{fang, mizuguchi09a} has reported evidence that this phase also shows up as an ``impurity" even in stoichiometric FeSe and $\textrm{Fe}_{0.8}\textrm{Co}_{0.2}\textrm{Se}$, suggesting that the hexagonal phase should be present even for Co concentrations of less than 25\%. Gathering all this evidence, we arrive at the following structural picture: At room temperature, FeSe is known to be antiferromagnetic (AFM) and possesses a tetragonal structure; while Co- and Ni-substituted FeSe exhibit superconductivity in a narrow band of substitution, namely, less than 15\% and 10\%, respectively, \cite{mizuguchi09a,liu1,shipra} where the tetragonal structure still dominates but starts to compete with the hexagonal one. CoSe, which is nonmagnetic (NM) at room temperature, has a NiAs-type hexagonal structure. Thus, as Co content increases, the tetragonal phase vanishes. The tendency points to a crossover that occurs at 38\% Co substitution from where the hexagonal phase dominates the system. This fact might help us understand why superconductivity in Co-substituted FeSe is suppressed even for relatively low values of Co substitution of less than 15\%. For comparison, in the $\textrm{FeSe}_{1-x}\textrm{Te}_{x}$ system superconductivity exists for Te substitutions as large as 90\%, where no structural transition occurs despite that the lattice distorts and the size of the lattice increases in proportion to Te substitution.\cite{fang,mizuguchi09a,iperez14b} The crystallographic information for $\textrm{FeSe}_{1-x}\textrm{Te}_{x}$ also reveals traces of a hexagonal phase in stoichiometric FeSe that vanishes for Te concentrations of more than 50\%. S-substituted FeSe shows a similar behavior, eliminating the hexagonal phase also for S concentrations greater than 50\%. We therefore see that whereas the effect of Te and S on the crystalline properties of FeSe is to eliminate the hexagonal phase, the effect of Co in FeSe is to eliminate the tetragonal one. XRD analysis performed on samples of $\textrm{FeSe}_{1-x}\textrm{Te}_{x}$ doped with transition metals such as Cr, Mn, Co, Ni, Cu, and Zn (5\% doping level) suggests that the atomic ionic radii of dopants may be crucial in effectively doping the host system. Since the ionic radii of Cr and Mn are larger than those of Co, Ni or Cu, Cr and Mn are not effectively incorporated by the host system as the energy-dispersive X-ray measurements demonstrate.\cite{zhang11a} Hence, bearing in mind that Co donates one $3d$ electron to the system, we conclude that the addition of this extra electron along with the increase of the nuclear charge induce a transition in both the crystalline structure and the magnetic order which in turn destroys the superconducting state. In fact, we shall show below that the magnetic order can be related to the spin state of the system; information that can be extracted from our RIXS measurements.  

\subsection{X-ray spectroscopy measurements}
We measured the TM $L_{2,3}$ RIXS and non-resonant XES spectra for the $\textrm{Fe}_{1-x}\textrm{Co}_x\textrm{Se}$ system. The data were collected at the soft x-ray fluorescence endstation at Beamline 8.0.1 of the Advanced Light Source (ALS) at Lawrence Berkeley National Laboratory. The endstation has a Rowland circle geometry X-ray spectrometer with spherical gratings and an area-sensitive multichannel detector.\cite{jia} The instrumental resolving power (E/$\Delta E$) for XES spectra was approximately $10^3$. Emission spectra were normalized with respect to the same peak height of the TM $L_3$ peak. X-ray absorption spectroscopy (XAS) measurements were carried out to determine only the excitation energies for the RIXS measurements. These energies corresponded to the location of the $L_2$  and $L_3$ peaks, an energy between them, and an energy well above the $L_2$ threshold for the non-resonant XES. For the XAS measurements we used the surface-sensitive total electron yield (TEY) mode. All XAS spectra were normalized to the beam flux measured by a clean gold mesh. The instrumental resolving power for all XAS measurements was about $5\times10^3$. During the measurements the samples were placed in ultra high vacuum (better than $10^{-8}$ Torr) and measured at room temperature. Finally, for reference, we also measured the emission spectra of Fe and Co metals, FeO and CoO. 

\subsection{Calculation details}
Electronic structure calculations were performed using DFT with the full-potential linearized augmented plane-wave (LAPW) method as implemented in the WIEN2k code.\cite{blaha} For the exchange correlation potential we employed the generalized gradient approximation in the Perdew-Burke-Ernzerhof variant.\cite{perdew} We generated a $12 \times12\times7$ k-mesh to perform the Brillouin zone integrations and for the expansion of the basis set we chose $R^{min}_{MT} K_{max}=7$ (the product of the smallest of the atomic sphere radii $R_{MT}$ and the plane wave cutoff parameter $K_{max}$). The radii of the muffin-tin spheres for the atoms were chosen so that the neighbouring spheres were nearly touching. The values adopted were: $R_{\textrm{Fe}}=2.21$, $R_{\textrm{Se}}=1.96$ and $R_{\textrm{Co}}=2.00$. For the calculations we used the experimental values of the lattice parameters according to the results of Table \ref{fecosetable}. To simulate Co substitution for the stoichiometries with $x=0.25$, 0.50, and 0.75, we generated $2\times 2\times 1$ supercells. In these supercells, a substitution of two Fe atoms by two Co atoms represents a 25\% substitution. The space groups for these structures were, respectively: orthorhombic $Pmm2$, $Pmm2$, and tetragonal $P$-$4m2$. The supercell calculations were not optimized because the experimental values are well known from XRD.\cite{gresty,viennois1} As mentioned above, we used hexagonal and tetragonal structures, \cite{wyckoff} with space group $P6_3/mmc$ and $P4/nmm$, for CoSe and FeSe, respectively. In all cases energy convergence of 0.0001 Ryd, charge convergence of 0.001 e and cutoff between core and valence states of -6 Ryd were chosen.

Finally, in order to interpret our emission measurements, XES spectra were also calculated using the XSPEC package implemented in WIEN2k.\cite{schwarz} The package calculates the spectra based on the dipole allowed transitions which are then multiplied with a radial transition probability and the partial densities of states.

\section{Results and discussion}
Above we already established that the introduction of Co in FeSe induces a structural transition from tetragonal to hexagonal accompanied by a transition in the magnetic order. We now study the influence of the Co $3d$ states on the electronic properties of FeSe superconductor. Our tools are capable of exploring these matters and we will start by analyzing the partial density of states. Later, we move on to determine the spin state of our system (which can be correlated to the magnetic order) using the information provided by the TM $L_{2,3}$ RIXS spectra. By analyzing the non-resonant XES at the TM $ L_{2,3}$-edge, we then study the valence states and determine the importance of the electronic correlations. This is realized by contrasting our calculated XES spectra with the measurements and by comparig to well-known correlated system such as transition metal oxides, and, to some degree by presenting exact calculations of the Hubbard mean field approximation (HMFA). With these tools, we show that FeSe is slightly more correlated than both the rest of the series and the iron-pnictide superconductors.

\subsection{Density of States}
\begin{figure*}[tp]
\begin{center}
\includegraphics[width=10cm,angle=270]{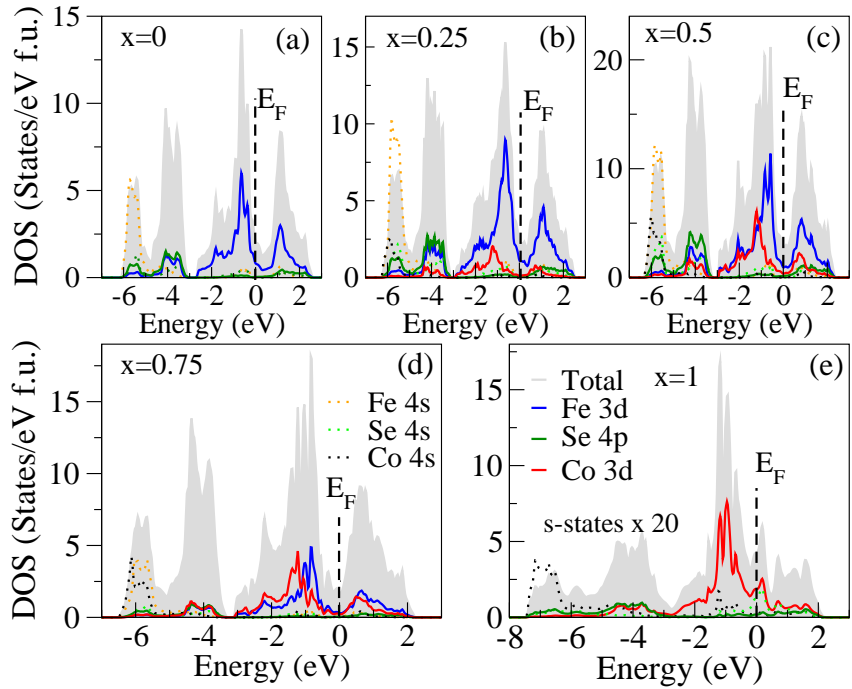}
\caption{(Color online) Density of states for $\textrm{Fe}_{1-x}\textrm{Co}_x\textrm{Se}$. The Fermi energy ($E_F$) is set at zero energy. For the sake of visualization the PDOS for Fe, Co and Se $s$-states are multiplied by a factor of 20 (dotted spectra).}
\label{dosfecose}
\end{center}
\end{figure*}

We show the calculated density of states (DOS) for the system $\textrm{Fe}_{1-x}\textrm{Co}_x\textrm{Se}$ in Fig. \ref{dosfecose}. The calculations for FeSe and CoSe are consistent with those reported in the literature.\cite{kurmaev,iperez14b,wadati,ikeda96a} First we notice that the valence band can be divided in two main regions. Near the Fermi level the DOS is amply dominated by a band of TM $3d$ states extending from $0.0$ eV to $-2.7$ eV. After a small gap of approximately 0.3 eV, a second band of hybridized TM $3d$ and Se $4p$ electrons extends from $-3.0$ eV to $-6.1$ eV. The states close to the Fermi energy for $0<x<1$ show mixed Fe and Co $3d$ character with the bulk of Fe $3d$ states dominating up to $x<0.75$. 
In a previous investigation, it was thought that for Se-deficient FeSe a reduction in the number of Fermi level $3d$ states may be important for suppressing superconductivity.\cite{kurmaev} In Table \ref{values} we show the values found for our system and compare them with the value for $\textrm{LaFeAs}(\textrm{O}_{1-x}\textrm{F}_x)$.
 \begin{table}[bp]
   \caption{Calculated total DOS at the Fermi level. As reference, we also added the value for the system $\textrm{LaFeAs}(\textrm{O}_{1-x}\textrm{F}_x$) taken from Singh \cite{singh}.}
  \centering 
  \begin{tabular}{c|c|c}
  \hline   \hline
Sample & Space group & N($E_F$) \\
&& \footnotesize{states/eV f.u.}
  \\ \hline   \hline
\footnotesize{FeSe} & \footnotesize{$P4/nmm$} & \footnotesize{1.04} \\
\footnotesize{$\textrm{Fe}_{0.75}\textrm{Co}_{0.25}\textrm{Se}$} &   \footnotesize{ $Pmm2$} & \footnotesize{1.08} \\
\footnotesize{$\textrm{Fe}_{0.5}\textrm{Co}_{0.5}\textrm{Se}$} & \footnotesize{$Pmm2$ } & \footnotesize{1.30} \\
\footnotesize{$\textrm{Fe}_{0.25}\textrm{Co}_{0.75}\textrm{Se}$} & \footnotesize{$P$-$4m2$} & \footnotesize{0.87} \\
\footnotesize{CoSe} & \footnotesize{$P6_3/mmc$} & \footnotesize{2.13} \\
\footnotesize{$\textrm{LaFeAs}(\textrm{O}_{1-x}\textrm{F}_x$)} & \footnotesize{$P4/nmm$} & \footnotesize{2.62} \\
   \hline   \hline
\end{tabular}
 \label{values}
\end{table}
Note that the values are comparable for $x<1$ and reach a maximum at $x=0.50$. Given that superconductivity in these materials occurs for Co substitution of at most 15\%, our results show that there may not be a direct correlation between the density of states at the Fermi level and the appearance of superconductivity. What is evident is that there is a relative shift of 0.7 eV between the center of mass of Fe and Co $3d$ states. This shift is expected for isovalent substitution and is consistent with the change of nuclear charge, implying that the extra Co electron resides at the Co site in agreement with impurity models.\cite{wadati,berlijn} If we now compare the DOS of our system with the DOS of the iron-pnictide superconductors, we discover substantial similitudes. \cite{kurmaev1,mcleod5} In particular, the major contribution comes from the TM $3d$ states in the neighbourhood of the Fermi level and a minor contribution from the chalcogen $s$ and $p$ states. Because the DOS of $\textrm{Fe}_{1-x}\textrm{Co}_x\textrm{Se}$ is so similar to those of other iron pnictides which are known to be weakly correlated, this suggests that $\textrm{Fe}_{1-x}\textrm{Co}_x\textrm{Se}$ can be viewed as a weakly correlated material as well. Indeed, in the following sections we will present experimental evidence supporting this view when we calculate the $L$-edge XES spectra for the TM species.

\subsection{Spin state and magnetic ordering}
RIXS is a powerful technique that is gaining wide recognition in condensed matter physics not only due to its element- and site-specific capabilities but also because its bulk sensitivity.\cite{kotani01a} When applied to TM compounds, this technique can probe the valence states of the TM and provide valuable information about the electronic structure. One of its versatile capabilities is that it can reveal information about spin state and magnetic ordering of a material. In this section we will focus on these properties.

\begin{figure*}[tp]
\begin{center}
\includegraphics[width=10cm,angle=270]{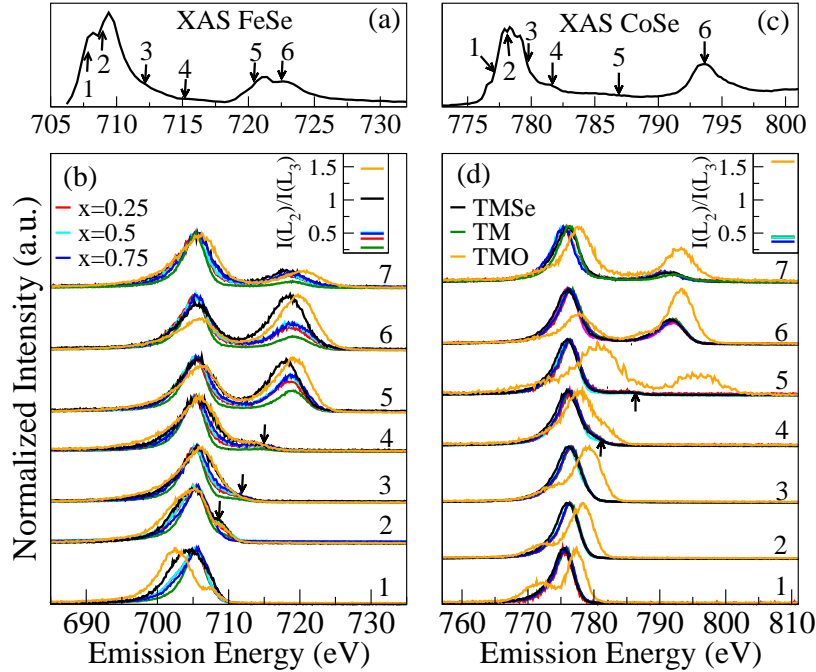}
\caption{(Color online) Fe (a) and Co (c) $2p$ XAS spectra for representative stoichiometries. Arrows in these spectra indicate the excitations energies used to collect the RIXS data.  Fe (b) and Co (d) $L_{2,3}$ RIXS spectra for the $\textrm{Fe}_{1-x}\textrm{Co}_x\textrm{Se}$ system along with Fe and Co metals (TM) and their monoxides (TMO). Arrows in the RIXS spectra indicate the appearance of small shoulders tracking the excitation energies. The non-resonant Fe and Co XES spectra were taken at energies of 740 eV and 820 eV, respectively, well above the $L_2$ threshold.  The insets show the resonant ratio for the resonant spectra at $L_2$ threshold (spectra No. 6).}
\label{fecoserixs}
\end{center}
\end{figure*}

In Fig. \ref{fecoserixs} we display the RIXS measurements of the TM $L$ edge for $\textrm{Fe}_{1-x}\textrm{Co}_x\textrm{Se}$. For reference we also include the RIXS spectra for Fe and Co metals and their respective monoxides. The top panels show the TM $2p$ XAS spectrum for FeSe and CoSe, respectively, where the arrows indicate the excitation energies (1-6) used to record the RIXS spectra. The spectra labelled with a number 7, correspond to the non-resonant emission taken at excitation energies of 740 eV and 820 eV, respectively.

To begin with this analysis let us first focus on the Fe $L$ RIXS spectra for $\textrm{Fe}_{1-x}\textrm{Co}_x\textrm{Se}$. The spectra number 7 show the two main fluorescence intensities resulting from the spin-orbital splitting which are centered at 705 eV and 718 eV and correspond to Fe $L_3$ and $L_2$ emission lines produced from the transitions $3d4s\to 2p_{3/2}$ and $3d4s\to 2p_{1/2}$, respectively. The spectra 2-4 displays small shoulders (marked with arrows) at energies between 708 eV and 715 eV. These energies track the excitation energies and, therefore, are caused by elastic scattering. In general we note that the spectra are featureless, resembling the spectrum of Fe metal, and their shape is independent of the excitation energy. It is also evident that variations in the Co content do not affect the overall shape of the Fe spectra, suggesting that there is no charge transfer between Fe and Co. These observations lend evidence for a delocalized character of the $3d$ electrons in these materials.\cite{schwarz,kurmaev3,gaoa} Likewise, the non-resonant XES spectra are quite similar to those reported in the Fe-pnictide superconductors.\cite{kurmaev1,kurmaev2,mcleod5,yang1} By contrast, the spectra for FeO show signs of constant energy loss features and of low-energy shoulders, and the spectral shape depends strongly on the excitation energy. These features are typical signatures of strong correlation effects such as charge transfer or $d$-$d$ excitations.\cite{prince}

Turning our attention to the Co $L$ RIXS spectra, we see that spectra 7 show the two main fluorescence bands $L_3$ and $L_2$ related to the spin-orbital splitting ($3d4s\to2p$ transition) which are located at approximately 776 eV and 792 eV, respectively. As before, the arrows in the spectra 4 and 5 indicate the appearance of small shoulders tracking the excitation energies which, again, can be attributed to elastic scattering. The analysis perfomed on the Co edge of $\textrm{Fe}_{1-x}\textrm{Co}_x\textrm{Se}$ reveals that Co in this system exhibits metallic character, since no multiplet features and no constant energy loss features are observed. For comparison, the spectra of CoO display a series of energy-loss structures which are identified as due to charge transfer and $d$-$d$ excitations.\cite{magnuson} These findings suggest that $\textrm{Fe}_{1-x}\textrm{Co}_x\textrm{Se}$ mainly behaves at most as a moderately correlated system with no strong bonds between TM and ligands.

Of course, one cannot overlook the fact that the Fe edge manifests evident differences in the ratio of the integrated intensities $L_2$ and $L_3$, $I(L_2)/I(L_3)$, when excited at the $L_2$ threshold (spectra 5 and 6) --- to avoid any confusion with the {\it non-resonant ratio} (NRR) from non-resonant XES (see below), it would be convenient to denote this new ratio as the {\it resonant ratio} (RR). Previous work reported similar behaviour in the RR of Fe compounds such as FeO, FeS$_2$ and $\textrm{Fe}\textrm{Se}_{1-x}\textrm{Te}_x$.\cite{iperez14b,prince} These variations are attributed to variations in the Coster-Kronig rate which in turn is linked to the spin state of the material. A rule for resonant emission at the $L_2$-edge developed by Prince et al. \cite{prince} states that the RR is higher for high spin ground states than for low spin ground states. According to this, the RR can be used to characterize the magnetic state of a material since the spin state is related to the magnetic ordering. Thus, under this method, magnetic $3d$ materials tend to exhibit high RRs whereas nonmagnetic materials show low ones. In the insets of Fig. \ref{fecoserixs}(b) and \ref{fecoserixs}(d) we show the RRs for the Fe and Co edges, respectively. It is well known that FeO is an antiferromagnetic material and good agreement with experiment is observed if one assumes that Fe is in a high spin state (S=2) with valency 2+ in cubic symmetry (and similarly for CoO). Prince et al. found that the RR for FeO is about 1.35 which is comparable to ours (RR$=1.45$). Here we find that the value for FeSe (RR$=1.1$) is closer to that of FeO than to that of Fe metal (RR$=0.3$). They also reported that FeS$_2$ is nonmagnetic, with a low spin state (S=0) and a valency of 2+ for Fe in octahedral symmetry. For this system they found a RR of 0.47 which is comparable to the value of Fe metal. The values for the rest of the series in $\textrm{Fe}_{1-x}\textrm{Co}_x\textrm{Se}$ are also below 0.47, indicating a low spin state (S=0). This is in agreement with the nonmagnetic character of CoSe. If we now turn our attention to the RR for the Co edge, we reaffirm also that all stoichiometries containing Co are in a low spin state which is consistent with the information obtained from the Fe edge. Very recent reports using polarized Raman-scattering techniques \cite{gnezdilov} demonstrate that FeSe undergoes spin fluctuations as function of increasing temperature; manifesting higher spin states at high temperatures. The estimations of the spin state at room temperature reveal that FeSe is in a high spin state (S=2). Remarkably, this is in qualitative agreement with our findings using the RIXS technique. For comparison, applying the same technique, we find that the system $\textrm{FeSe}_{1-x}\textrm{Te}_{x}$ shows several spin states as follows: S=0 for $x=0.50$; S=1 for $x=0.25$, 0.75, and 1.00; and S=2 for $x=0.00$. This corroborates that FeSe is in a high spin state. \cite{iperez14b}

\subsection{Strength of electronic correlations}
\begin{figure*}[tp]
\begin{center}
\includegraphics[width=10cm, angle=270]{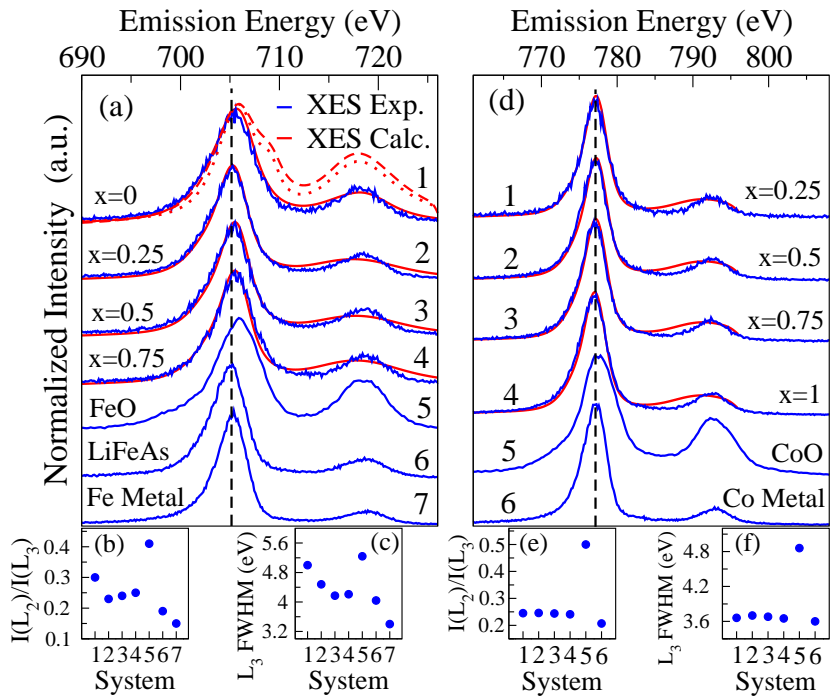} 
\caption{(Color online) Comparison of non-resonant XES spectra (solid blue). (a) Fe $L_{2,3}$ edge for $\textrm{Fe}_{1-x}\textrm{Co}_x\textrm{Se}$, FeO, LiFeAs and Fe metal. (b) Non-resonant ratio and (c) FWHM of $L_3$ for the systems in panel (a) indexed with numbers. (d) Co $L_{2,3}$ edge for $\textrm{Fe}_{1-x}\textrm{Co}_x\textrm{Se}$, CoO and Co metal. (e) and (f) are similar to (b) and (c) but for the Co edge taken from (d). The calculated XES spectra for TM edges for $\textrm{Fe}_{1-x}\textrm{Co}_x\textrm{Se}$ are shown in red in (a) and (d), respectively. The LDA+U calculations for FeSe with Coulumb parameter $U=0.5$ Ry (dotted) and $U=1.5$ Ry (dashed) are also shown in (a).}
\label{fecosexes}
\end{center}
\end{figure*}
The non-resonant XES spectra are interpreted as giving direct information on the occupied DOS. Figure \ref{fecosexes} displays the measurements (solid blue spectra) for the non-resonant TM $L_{2,3}$ XES of $\textrm{Fe}_{1-x}\textrm{Co}_x\textrm{Se}$ in comparison to those of Fe and Co metals, FeO, LiFeAs, and CoO. To interpret our measurements, we have also included the calculated XES spectra (in red). These spectra were Lorentzian- and Gaussian-broadened to account for lifetime and instrumental broadening and manually shifted in energy to match the measured spectra. We can see that the calculations are quite consistent with the measurements although, due to broadening, no subtleties are revealed. Despite this, the simple line shape in the non-resonant XES reveals that there is no sign of a lower Hubbard band or any other complexities that arise in strongly correlated systems. The lack of charge transfer or $d$-$d$ excitation features in the RIXS spectra suggests that the TM $3d$ states do not form strong bonds with chalcogen states outside the $3d$ band. It also supports the DFT picture that $\textrm{Fe}_{1-x}\textrm{Co}_x\textrm{Se}$ is at most a moderately correlated system. 

The Fe measurements for $\textrm{Fe}_{1-x}\textrm{Co}_x\textrm{Se}$ are quite similar in both energy and shape to those of Fe metal and LiFeAs. This shows that the distribution of Fe $3d$ states in our samples resembles that of Fe metal and reflects the itinerant character of the system. In general we see that the increase of $x$ does not influence the shape of the spectra corroborating our previous assumption that there is no considerable Fe-Co charge transfer. In contrast, notice that the spectrum for FeO is shifted about 1 eV to the right if compared to the other systems and shows a shoulder at the low energy of the main fluorescence peak which is characteristic of most TM oxides.\cite{prince} This indicates a formal change of Fe valency from 0 in Fe metal to 2+ in FeO. In order to give a quantitative assessment of these observations, we have computed the ratio of the integrals under the peak intensities or the NRR. This quantity provides valuable information related to the population of $2p_{1/2}$ and $2p_{3/2}$ energy levels, respectively. In the one-electron approximation the NRR should equal 0.5. Nevertheless, in metals, Coster-Kronig (C-K) transitions reduce the emission probability for the $L_2$ peak; thus revealing information of the electronic correlations of the system and its metallicity.\cite{kurmaev3} As such, this method has been used to characterize the correlation strength in iron pnictides and other transition metal compounds.\cite{kurmaev,iperez14b,mcleod5,mcleod} The results of the NRR for all systems are shown in Fig. \ref{fecosexes}(b). The values for $\textrm{Fe}_{1-x}\textrm{Co}_x\textrm{Se}$ are between Fe metal and FeO, indicating that the $3d$ electrons are more localized in these systems than in Fe-pnictide superconductors. The value of FeSe is closer to FeO than to Fe metal whereas the value for the rest of the series is closer to Fe metal, suggesting again that $\textrm{Fe}_{1-x}\textrm{Co}_x\textrm{Se}$ can be considered at most as a moderate correlated system. This is an important finding if we keep in mind that in most Fe-pnictide superconductors the $3d$ electrons are mainly itinerant.\cite{kurmaev2,yang1} The metallic character of these materials is represented by LiFeAs whose value is quite close to the value of Fe metal. 

On the other hand, the bandwidth of $L_3$ could be associated with the electronic distribution near the Fermi energy. Figure \ref{fecosexes}(c) shows the bandwidth of the $L_3$ edge taken from the spectra in (a). The bandwidth of $\textrm{Fe}_{1-x}\textrm{Co}_x\textrm{Se}$ is smaller than that of FeO but, again, the value for FeSe is closer to FeO than to Fe metal. This illustrates that the bulk of the $3d$ electrons for $x>0$ is localized in a narrow band near the Fermi energy as found in our DOS calculations, unlike the case of the FeO whose $3d$ states are spread over a wider region relatively far from the Fermi level. We also note that the bandwidth decreases as $x$ decreases and reaches a minimum at $x=0.5$ and then slightly increases for $x=0.75$. For some Fe-pnictides and metallic compounds, the change in the Fe-Fe distance is linked to the Fe $3d$ bandwidth. The reports show that bandwidth decreases with increasing Fe-Fe distance.\cite{kurmaev2} In our case, however, we did not detect any clear trend that could be correlated to the crystal properties.

Analyzing now the Co $L$ edge (shown in Fig. \ref{fecosexes}(d)), we can tell a similar story as in the case of the Fe edge. Neither low-energy shoulders nor multiplet satellites in any of our stoichiometries are apparent. Instead, we observe a great similarity with Co metal, indicating that the majority of Co $3d$ states is concentrated near the Fermi level. The values obtained for the non-resonant ratio (Fig. \ref{fecosexes}(e)) as well as those for the FWHM of $L_3$ (Fig. \ref{fecosexes}(f)) of $\textrm{Fe}_{1-x}\textrm{Co}_x\textrm{Se}$ reinforce its metallic character. In contrast, the Co $L$ edge of CoO shows a small shoulder around 770 eV and the spectrum is slightly shifted to the right. Furthermore, its NNR and FWHM of $L_3$ are much greater than in the other systems.\cite{magnuson} These differences suggest again that $\textrm{Fe}_{1-x}\textrm{Co}_x\textrm{Se}$ is at most a moderately correlated material.

The fact that in FeSe the C-K transitions are more suppressed than in the rest of the series suggests that the Fe $3d$ states are more localized and, indeed, we show evidence that FeSe is slightly more correlated than both the rest of the series and the iron-pnictide superconductors. In this respect, other authors have reported theoretical and experimental evidence for strong correlation effects in FeSe. \cite{miyake10a,aichhorn1,yamasaki10a} In order to further estimate the correlations in FeSe, we performed LDA+U calculations within the context of the HMFA. The results show that for values of the Coulumb parameter $U$ ($<1.5$ Ry) and of the Hund coupling $J$ ($<0.5$ Ry), the total DOS drastically changes and new shoulders in the XES spectra appear. Figure \ref{fecosexes} (a) shows the calculated Fe XES spectra for FeSe with $U=0.5$ Ry (dotted) and $U=1.5$ Ry (dashed), respectively, both with $J=0$ Ry. The evident discrepancy with experiment indicates that $U$ must be very small and that our initial theoretical considerations suffice to reproduce the observations.

Finally, we would like to mention that some researchers\cite{anisimov09a,johnston10a} consider the ratio of the on-site Coulumb repulsion $U$ to the conduction electron bandwidth W ($U/W$) as a parameter to quantify the correlation strength within the context of the LDA++, the LDA+U and some other approaches developed for strongly correlated systems.\cite{anisimov91a} Within these contexts one can distinguish three regimes, namely: $U/W>1$ for strong correlations, $U/W\approx 1$ for intermediate correlations and $U/W<1$ for weak correlations. As shown in Fig. \ref{fecosexes}(c) the FWHM of the Fe $L_3$ XES decreases about 1.5 eV from $x = 0$ to $x = 0.75$. Thus, unless $U$ decreases by a greater rate, the correlation strength might actually increase. However, our results clearly show that this is not the case, since the calculated DOS in Fig. \ref{dosfecose} shows that $W$ is actually increasing as $x$ approaches 1, and the agreement between calculated and measured XES spectra in Fig. \ref{fecosexes}(a) shows that the FWHM decrease is due rather to a greater weighting of Fe $3d$ states in the main feature rather than a shrinking valence band. 

\section{Conclusions}

We study the effect of Co substitution on the Fe valence states of FeSe supercondutor using X-ray difracction, soft X-ray spectroscopy and density functional theory. According to our findings, Fe-Se bonds in FeSe seem to be weakened by the introduction of Co into the system, transforming FeSe from a moderately correlated system to a weakly correlated one. In this respect we also find that the $\textrm{Fe}_{1-x}\textrm{Co}_x\textrm{Se}$ system is more correlated than the iron-pnictides. The spin state was also estimated by analyzing the RR. We find that FeSe has the highest spin state (S=2) whereas the rest of the series has a low spin state (S=0), implying a decrease in the magnetic ordering from magnetic to nonmagnetic (this appears to be in agreement with impurity models find in the literature\cite{liu3,berlijn}). Along with the change in the magnetic order, we also find an associated change in the crystal structure from tetragonal to hexagonal, not detected before due to the relatively small Co concentrations. These findings strongly suggest that the crystal structure plays a key role for the interplay between magnetism and superconductivity in iron-chalcogenide materials.

Lastly, we point out that, to the best of our knowledge, this is the first report studying the $\textrm{Fe}_{1-x}\textrm{Co}_x\textrm{Se}$ system using X-ray spectroscopy. 

\section*{Acknowledgements}
The authors gratefully acknowledge support from the Natural Sciences and Engineering Research Council of Canada (NSERC) and the Canada Research Chair program. This work was done with partial support from CONACYT M\'exico under grant 186142 and from Programa de Apoyo a Proyectos de Investigaci\'on e Innovaci\'on Tecnol\'ogica (PAPITT), UNAM under project IN115410. The Advanced Light Source is supported by the Director, Office of Science, Office of Basic Energy Sciences, of the U. S. Department of Energy under Contract No. DE-AC02-05CH11231. The computational part of this research was enabled by the use of computing resources provided by WestGrid and Compute/Calcul Canada. We are indebted to the anonymous reviewers for helpful comments and suggestions.

\end{document}